\documentclass[amsmath,pre]{revtex4}

\usepackage{graphicx}
\usepackage{dcolumn}
\usepackage{subfigure}
\usepackage{bm}

\begin{document}


\title{Discrete element simulations of stress distributions in silos:
crossover from two to three dimensions}

\author{James W. Landry}
 \email{jwlandr@sandia.gov}
\author{Gary S. Grest}
\author{Steven J. Plimpton}
 \affiliation{Sandia National Laboratories, Albuquerque, New Mexico 87185}

\date{\today}

\begin{abstract}
The transition from two-dimensional ($2D$) to three-dimensional ($3D$)
granular packings is studied using large-scale discrete element computer
simulations.  We focus on vertical stress profiles and examine how they
change with dimensionality.  We compare results for
packings in $2D$, quasi-$2D$ packings between flat plates, and $3D$
packings.  Analysis of these packings suggests that the Janssen theory
does not fully describe these packings, especially at the top of the
piles, where a hydrostatic-like region of vertical stress is visible in
all cases.  We find that the interior of the packing is far from
incipient failure, while in general, the forces at the walls are close
to incipient failure.
\end{abstract}

\maketitle

\section{\label{sec:introduction}Introduction}

The stress within a silo packed with granular material has long been of
interest in the engineering~\cite{Janssen1895,Nedderman1992} and
physics~\cite{JaegerOct1996} communities.  As early as 1895, Janssen, in
his pioneering work, constructed a model to describe the vertical stress
in a silo~\cite{Janssen1895}.  Treating the granular pack as a
continuous medium where a fraction $\kappa$ of vertical stress is
converted to horizontal stress, Janssen was able to derive a simple
functional form for the vertical stress.  One main assumption of this
model is that the forces of friction between particles and walls are at
the Coulomb failure criterion: $F_t = \mu_w F_n$, where $F_t$ is the
magnitude of the tangential friction force, $F_n$ is the normal force at
the wall, and $\mu_w$ is the coefficient of friction for particle-wall
contacts.  This assumption is also known as incipient failure.  This
theory qualitatively describes the crossover to a depth-independent
vertical stress, although quantitative discrepancies between the Janssen
theory and experiment have been observed.  The assumption of incipient
failure is believed to be the source of the discrepancies, but this has
never been conclusively proven due to the difficulty of experimental
measurements of internal stress.  Numerous improvements have been added
to the theory over time, but none of these have gained wide
acceptance~\cite{Nedderman1992}.

Recently, well-controlled experiments have been carried out on granular
packs in silos to test the suitability of Janssen's theory in ideal
conditions~\cite{VanelMay1999,VanelFeb2000}.  Vanel and
Cl\'{e}ment~\cite{VanelFeb2000} constructed their initial packings by
tapping a loose poured packing to increase its density.  The base of the
packing was then slowly moved downward to more fully mobilize the
grains, and then the packing was allowed to settle.  They measured the
apparent mass at the bottom of the silo as a function of the filling
mass of the silo.  Their experiment did not follow the Janssen form, and
they found the best agreement with a phenomenological model, which
contains elements of Janssen's original analysis.  We describe this
model in more detail in Sec III.  Other more recent experimental studies
in which the side walls were dragged upward to fully mobilize the
packing~\cite{OvarlezJun2003,BerthoApr2003,OvarlezSep2003} have found
agreement with the Janssen form.

A variety of discrete element and continuous finite element simulation
methods have been developed to describe the stresses in a silo due to
the difficulty of experimental verification.  Most discrete element
simulations thus far have been confined to two dimensional ($2D$)
systems.  However, there is wide disagreement within the community on
the predictive power of these models and the proper approach to take for
accurate simulation~\cite{MassonApr2000, HolstJan1999a, HolstJan1999b,
SanadOct2001}.  Those simulations that are carried out in three
dimensions ($3D$) usually do so via finite-element methods that yield
little information on the internal structure or forces in granular
packs~\cite{ChenOct2001,GuinesOct2001}.  Some $3D$ discrete element
simulations~\cite{MakseMay2000,SilbertMar2002} have been performed, but
these use periodic boundary conditions in the two directions
perpendicular to gravity.  Though these studies provide useful
information on the internal structure of these packings, they can give
no information on vertical stresses or forces induced by the walls of
the container.  We have recently carried out $3D$ discrete element simulations
of stress in granular materials in cylinders~\cite{LandryApr2003}.  Here
we compares simulations in $2D$ and $3D$ and study the crossover from
two to three dimensions.

We present large-scale discrete element simulations of granular
packings in $2D$, quasi-$2D$, and $3D$ containers (silos).  Our aim is
to understand the vertical stress profiles of these granular packings,
particularly the crossover from two to three dimensions, which has
largely been unexplored.  Both monodisperse and polydisperse particle
systems are studied.  We explicitly test the suitability of the Janssen
theory to the vertical stress profiles produced by our simulations and
test the validity of the Janssen theory's assumptions.  We find that the
interior of the packing is far from incipient failure, while in general,
the forces at the walls are close to incipient failure.  In all cases,
there is a region at the top of the pile where the particle-wall forces
are far from incipient failure and the Janssen form is not observed.

Section II discusses the simulation technique and the method used to
generate the packings.  Section III presents the vertical stress
profiles.  We discuss their characteristics and compare our results to
the classical theory of Janssen as well as two modified forms of the
Janssen analysis.  In Section IV we show how forces in the packings are
not at incipient failure at the top and in the bulk of the packing and
relate this to the failure of the Janssen analysis to fully explain our
packing.  Finally, we conclude and summarize the work in Section V.

\section{\label{sec:simulation}Simulation Method}

We present discrete element simulations in two and three dimensions of
model systems of $N$ mono and polydispersed spheres of fixed density
$\rho$.  The mass of a given particle is $m_i = \frac{\rho}{6} \pi
d_i^3$, where $d_i$ is the diameter of a given particle.  Particle
diameters are uniformly distributed between $\bar{d} \pm \Delta$.  For
this study, we use $\Delta/\bar{d} = 0$, $0.1$, and $0.5$.  $N$ varies
from 10,000 to 50,000 particles.  The system is constrained by either an
open rectangular box or an open cylinder with its axis along the
vertical $z$ direction.  The box has length $L$, width $W$, and height
$H$, centered at $x=y=0$ and bounded below with a flat base at $z=0$.
The cylinder has radius $R$ and is centered on $x=y=0$.  It is bounded
below with a flat base at $z=0$.  In some cases, a layer of
randomly-arranged immobilized particles approximately $2d$ high were
used to provide a rough base.  We vary the length $L$ and width $W$ of
the box and the radius $R$ of the cylinder.  This work builds on
previous 3D simulations of packings in a cylinder~\cite{LandryApr2003}.

The spheres interact only on contact through a spring-dashpot
interaction in the normal and tangential directions to their lines of
centers.  Contacting spheres $i$ and $j$ positioned at $\mathbf{r}_i$
and $\mathbf{r}_j$ experience a relative normal compression $\delta =
|\mathbf{r}_{ij} - d|$, where $\mathbf{r}_{ij} = \mathbf{r}_i -
\mathbf{r}_j$, which results in a force~\cite{Cundall1979}
\begin{equation}
\mathbf{F}_{ij} = \mathbf{F}_n + \mathbf{F}_t.
\end{equation}
The normal and tangential contact forces are given by 
\begin{equation}
\mathbf{F}_{n} = f(\delta/d) (k_n \delta {\mathbf n}_{ij} - m_{eff} \gamma_n \mathbf{v}_n)
\end{equation}
\begin{equation}
\mathbf{F}_{t} = f(\delta/d) (-k_t \mathbf{\Delta s}_t - m_{eff} \gamma_t \mathbf{v}_t)
\end{equation}
where $\mathbf{n}_{ij} = \mathbf{r}_{ij}/r_{ij}$, with $r_{ij} =
|\mathbf{r}_{ij}|$. $\mathbf{v}_n$ and $\mathbf{v}_t$ are the normal and
tangential components of the relative surface velocity, and $k_{n,t}$
and $\gamma_{n,t}$ are elastic and viscoelastic constants,
respectively. $m_{eff} = \frac{m_i m_j}{m_i + m_j}$ for interactions
between particles $i$ and $j$, and $m_{eff} = m_i$ for interactions
between particle $i$ and a wall, where the mass of the wall is assumed
to be infinite.  $f(x) = 1$ for Hookean (linear) contacts, while for
Hertzian contacts $f(x) = x^{D/2 - 1}$ --- the two force models differing
only in $3D$.  $\mathbf{\Delta s}_t$ is the elastic tangential
displacement between spheres, obtained by integrating tangential
relative velocities during elastic deformation for the lifetime of the
contact.  The magnitude of $\mathbf{\Delta s}_t$ is truncated as
necessary to satisfy a local Coulomb yield criterion $F_t \le \mu F_n$,
where $F_t \equiv |\mathbf{F}_t|$ and $F_n \equiv |\mathbf{F}_n|$ and
$\mu$ is the particle-particle friction coefficient.  Frictionless
spheres correspond to $\mu = 0$.  Particle-wall interactions are treated
identically, though the particle-wall friction coefficient $\mu_w$ is set
independently.  A more detailed description of the model is available
elsewhere~\cite{SilbertOct2001}.  We also tested an alternate form of
the $F_t$ force due to Haff and Werner~\cite{Haff1986} that does not
have history effects.  This model produces a hydrostatic stress profile
in all cases.  Without the relative transverse displacement in the force
model, $F_t = 0$ in the quasi-static limit and the walls cannot support
stress.

We present all the physical quantities in our simulations in terms of
$\bar{d}$, the average diameter; $\bar{m}$ the mass of a particle with
average diameter $\bar{d}$; and $g$, the force of gravity.  Most of
these simulations were run with $k_n = 2 \times 10^5 \bar{m}g/\bar{d}$,
$k_t = \frac{2}{7} k_n$, and $\gamma_n = 50\sqrt{g/\bar{d}}$.  For
Hookean springs we set $\gamma_t = 0$.  In this case, the coefficient of
restitution for $\Delta = 0$ is $\epsilon = 0.8$.  For Hertzian springs,
$\gamma_t = \gamma_n$.  For Hertzian springs in $D=3$, $\epsilon
\rightarrow 0$ as $v \rightarrow 0$, so that it takes much longer for
energy to drain out of the pack and for the kinetic energy per particle
to decrease to sufficiently low levels.  For this reason, it is much more
computationally expensive to use Hertzian springs to form packings. The
convenient time unit is $\tau = \sqrt{\bar{d}/g}$, the time it takes a
particle to fall its radius from rest under gravity.  For this set of
parameters, the time step $\delta t = 10^{-4} \tau$.  In most
simulations, the particle-particle friction and particle-wall friction
are the same, $\mu = \mu_w = 0.5$.

The two-dimensional ($2D$) simulations were performed by fixing $y=0$
for all particles and removing any velocity or acceleration components
in the $y$ direction.  Some polydispersity ($\Delta/\bar{d} > 0$) is
required to prevent crystallization in $2D$, so all $2D$ simulations
were run with a polydispersity of $\Delta/\bar{d} = 0.1$.

All of our results will be given in dimensionless units based on
$\bar{m}$, $\bar{d}$, and $g$.  Physical experiments often use glass
spheres of $\bar{d} = 100 \mu m$ with $\rho = 2 \times 10^3 kg/m^3$.  In
this case, the physical elastic constant would be $k_{glass} \sim
10^{10} \bar{m}g/\bar{d}$.  A spring constant this high would be
prohibitively computationally expensive, because the time step must have
the form $\delta t \propto k^{-\frac{1}{2}}$ for collisions to be
modeled effectively.  We have found that increasing $k$ in our
simulations does not appreciatively change the physical
results~\cite{SilbertOct2001}.

We generate our packings by mimicking the pouring of particles at a
fixed height $Z$ into the container.  For computational efficiency
a group of $M$ particles is added to the simulation on a single
time step.  This is done by inserting the $M$ particles at
non-overlapping positions within a thin cylindrical region of radius
$R-\bar{d}$ with height $Z-\bar{d} < z < Z$.  The $x$, $y$, and $z$
coordinates of the particles are chosen randomly within this insertion
region.  The height of insertion $z$ determines the initial z-velocity
$v_z$ of the particle --- $v_z$ is set to the value it would have after
falling from a height $Z$.  After a time $\sqrt{2} \tau$, another group
of $M$ particles is inserted.  This methodology generates a steady stream
of particles, as if they were poured continuously from a hopper (see
Figure 1).  The rate of pouring is controlled by setting $M$ to
correspond to a desired volume fraction of particles within the
insertion region.  For example, for an initial volume fraction of
$\phi_i = 0.13$ in a cylinder of $R=10d$, the pouring rate is $\approx
45$ particles/$\tau$.  This method is similar to the homogeneous
``raining'' methods used in experiments~\cite{VanelNov1999}.

The simulations were run until the kinetic energy per particle was less
than $10^{-8} \bar{m}g\bar{d}$.  The resultant packing is considered
quiescent and used for further analysis~\cite{SilbertMar2002}.  In $3D$,
Hertzian packings have difficulty reaching this criteria due to
long-lived pressure waves that traverse the entire pile.  In these
packings, the kinetic energy per particle tends to stabilize at
approximately $10^{-3} \bar{m}g\bar{d}$ for very long times.  By adding
a viscous damping term proportional to $v$ after the pack has stabilized
at $10^{-3} \bar{m}g\bar{d}$, we were able to reduce the kinetic energy
per particle of these systems to less than $10^{-8} \bar{m}g\bar{d}$.  A
comparison of the stress profiles before and after the application of
this viscous damping term showed no change.  Although a change of
kinetic energy of this order has a small effect on the coordination
number of the packing, it does not have any effect on the vertical
stress.  All vertical stress profiles for Hertzian packings presented in
the following sections were generated with this technique.

These simulations were performed on a parallel cluster computer built
with DEC Alpha processors and Myrinet interconnects using a parallel
simulation code optimized for short-range interactions
\cite{SilbertOct2001,PlimptonMar1995}.  A typical simulation to pour
$50,000$ particles into a 3D cylindrical silo requires $5 \times 10^6$
time steps, which requires roughly $40$ CPU hours on $50$ processors.

\begin{figure*}
\includegraphics[width=1.0in]{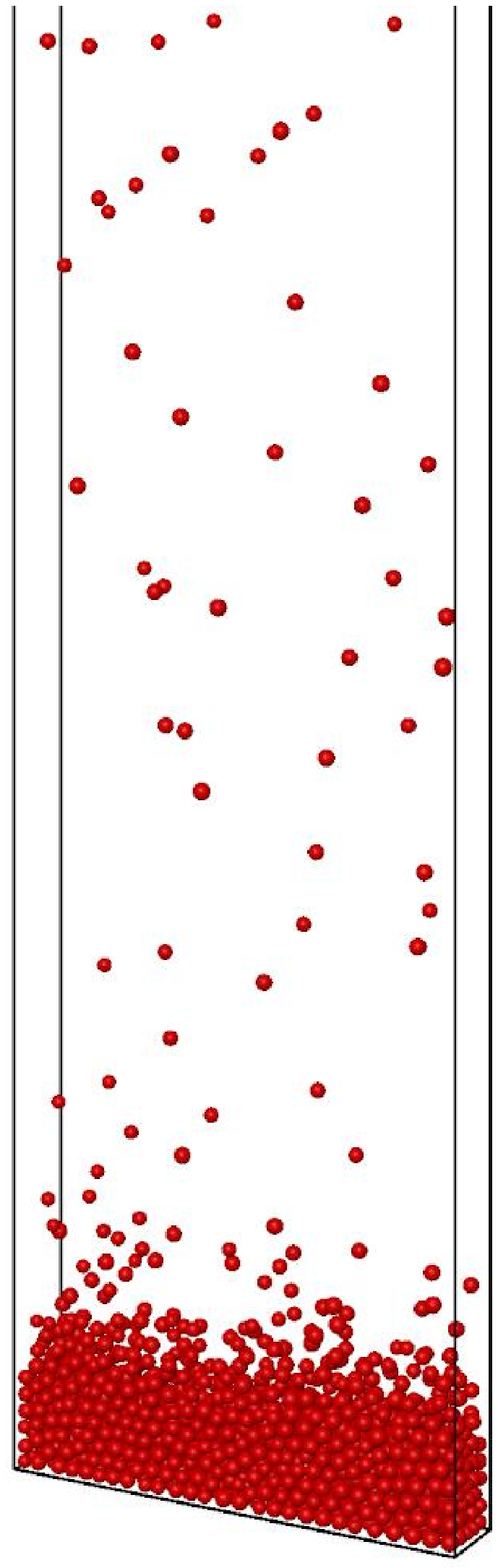}
\hspace{0.5in}
\includegraphics[width=1.0in]{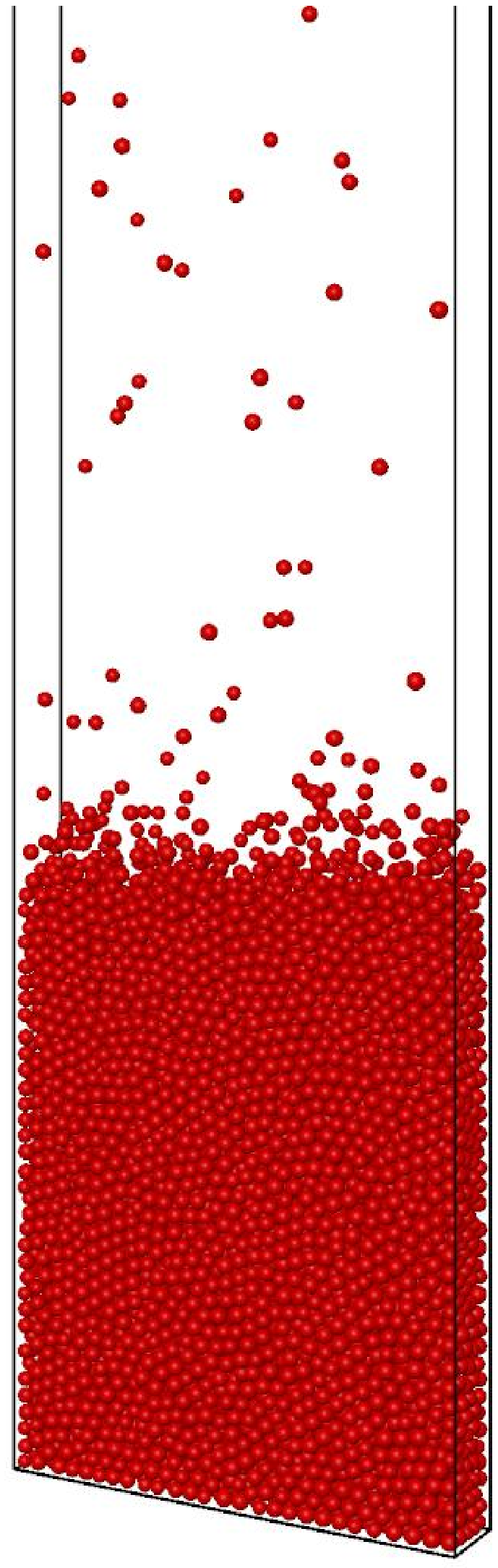}
\hspace{0.5in}
\includegraphics[width=1.0in]{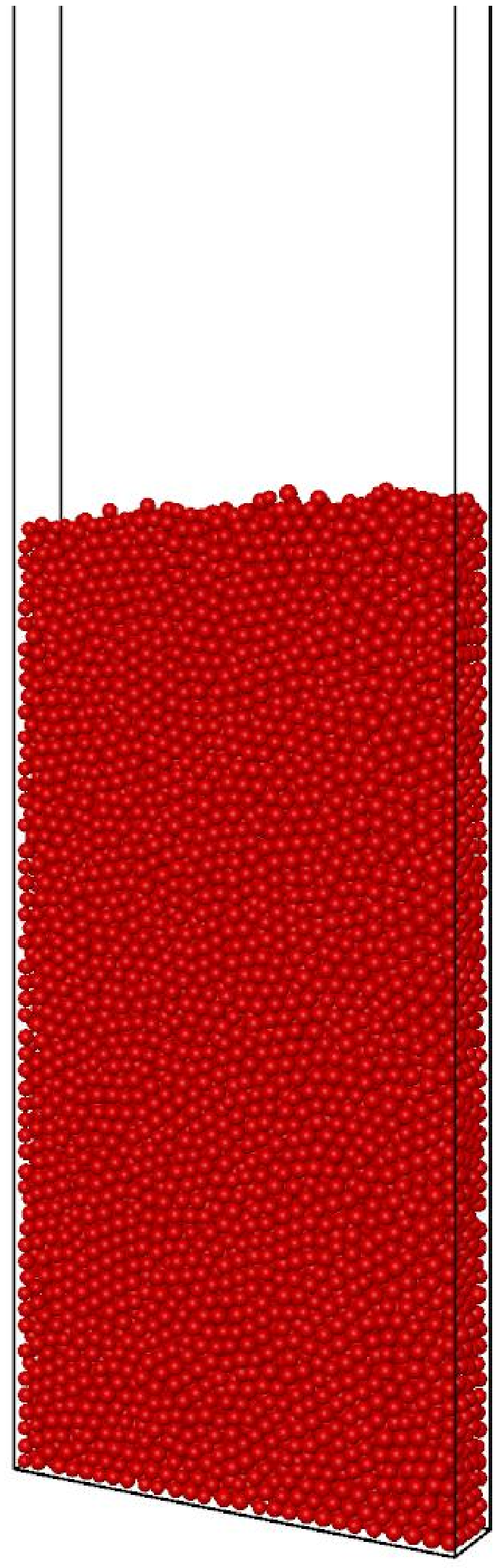}
\caption{\label{fig:pourmovie1} Formation of a packing of $N = 10,000$
spheres in a rectangular container of length $30\bar{d}$ and width
$5\bar{d}$.  The black lines show the container walls.  The packing is
constructed by pouring from a height of $401\bar{d}$.  The
configurations shown represent early times, late times, and the final
packing.}
\end{figure*}

We show in Figure~\ref{fig:pourmovie1} how a typical packing is created.
The figure shows three snapshots in time as particles are poured into a
quasi-$2D$ system of length $30\bar{d}$ and width $5\bar{d}$.

\begin{figure}
\includegraphics[width=1.7in]{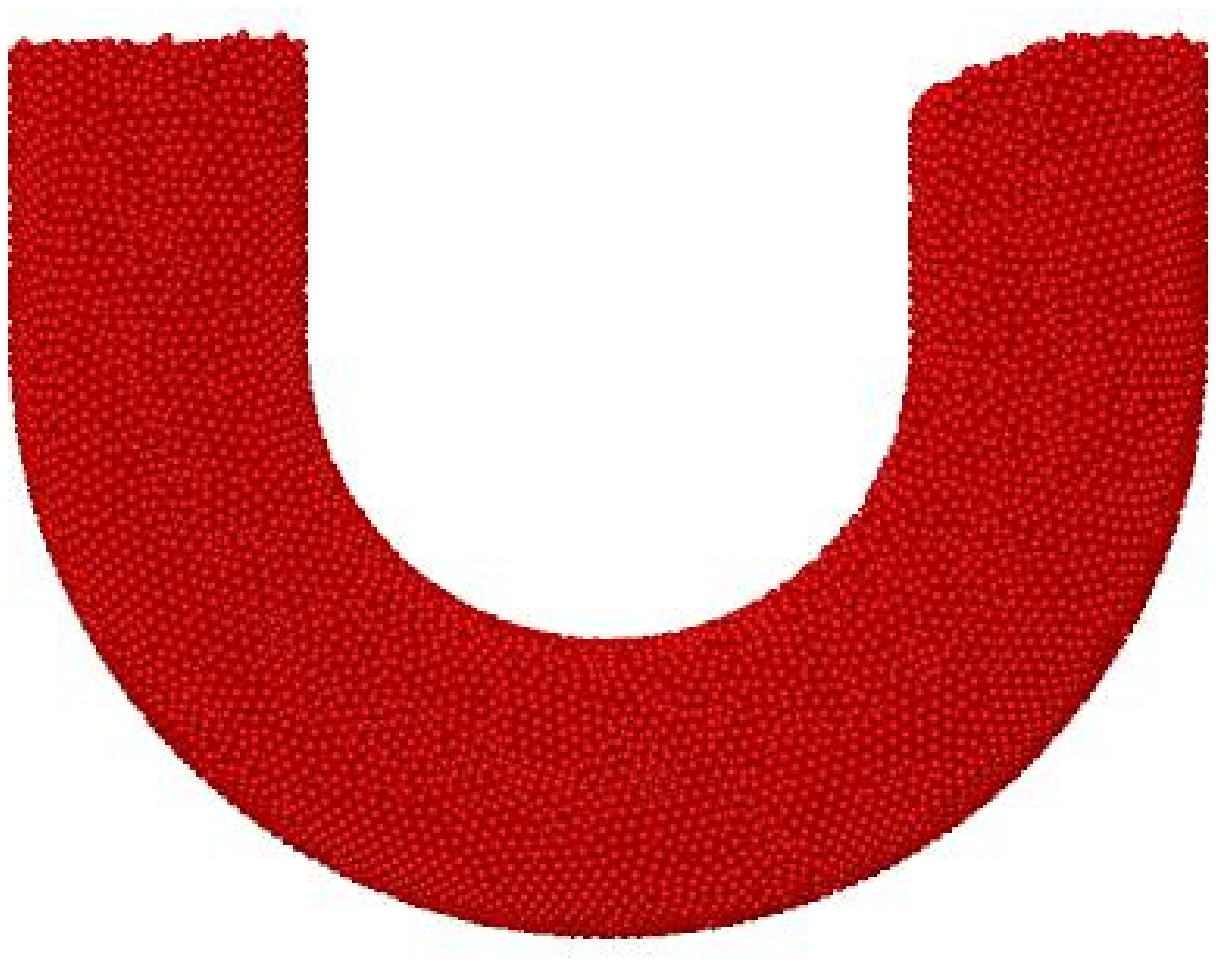}
\hspace{0.5in}
\includegraphics[width=1.7in]{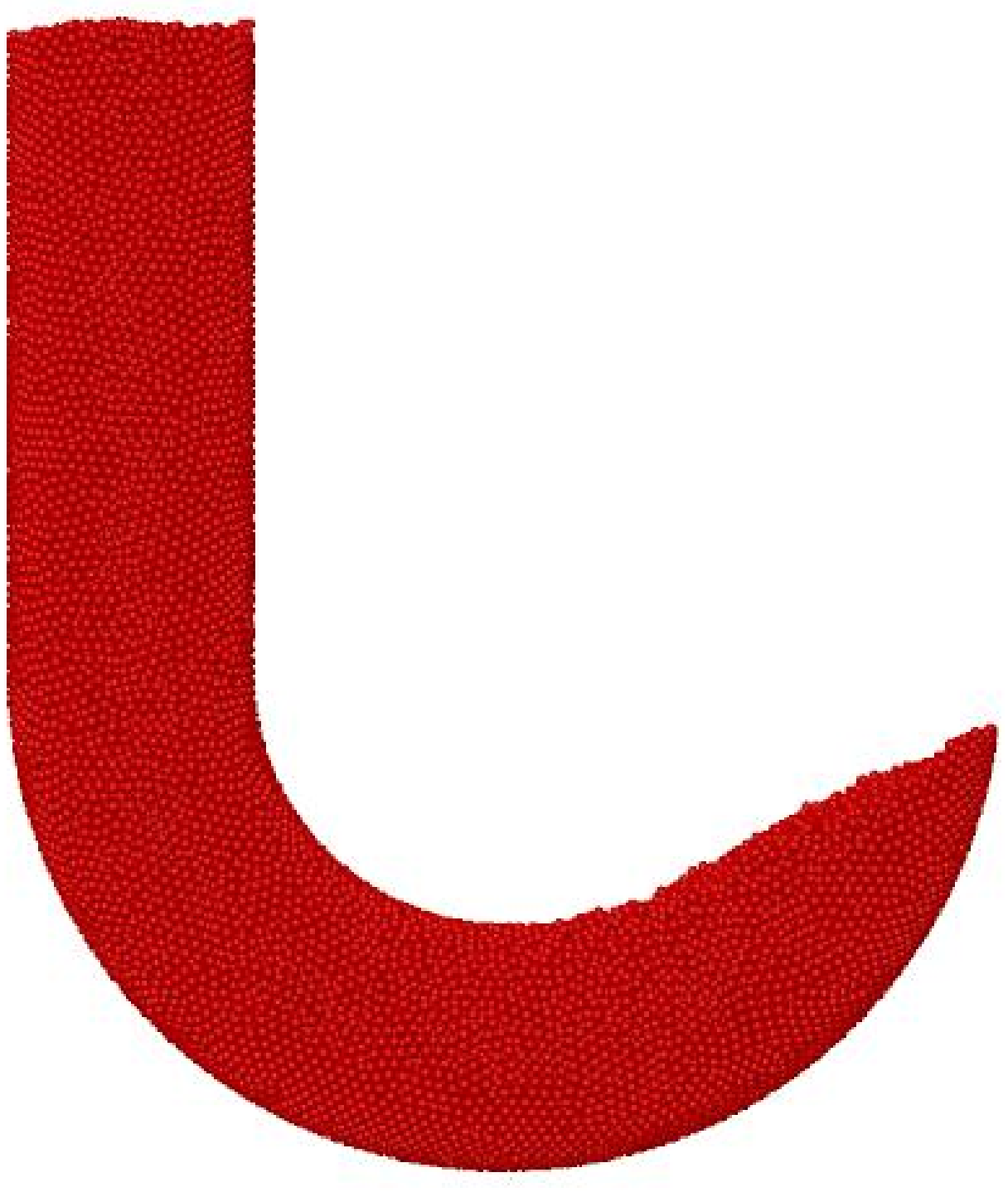}
\hspace{0.5in}
\includegraphics[width=1.7in]{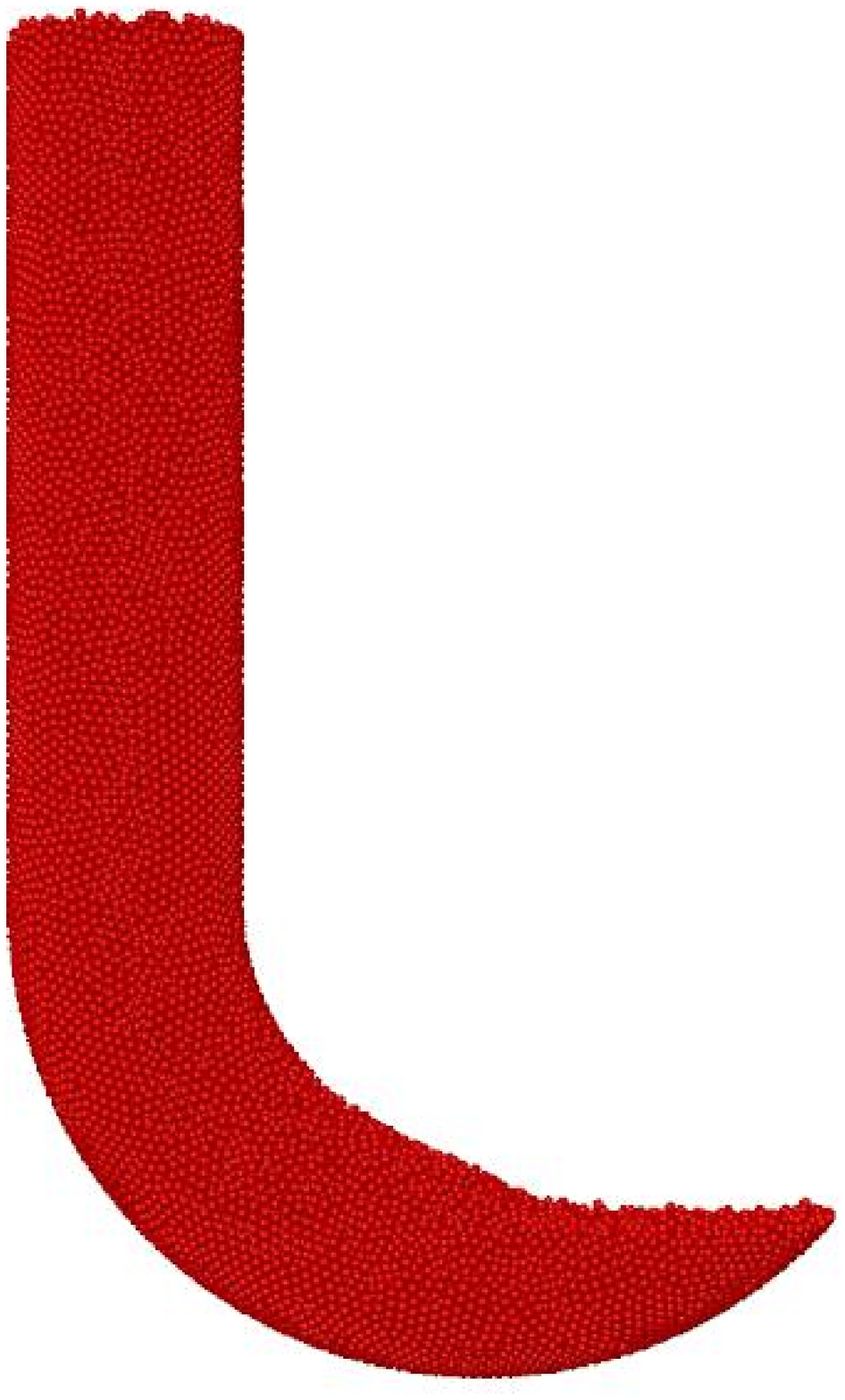}
\caption{\label{fig:utube} Packings of $N=10000$ particles in U-Tubes
with different $\mu_w$.  Particles were poured into the left-hand side
and allowed to settle into packings.  a) $\mu_w = 0.0$ and no stress is
carried by the walls.  b) $\mu_w = 0.1$ and the walls support slight
stresses and there is a height difference between the left and right
tubes.  c)$\mu_w = 0.5$ and there is a great difference between the left
and right tubes.}
\end{figure}

The strong effect of wall friction on the capacity of a silo to support
stress as found experimentally is shown in Figure~\ref{fig:utube} where
the particle-wall friction $\mu_w$ is varied, while keeping the
particle-particle friction constant at $\mu = 0.5$.  The geometry of the
system is a three-dimensional U-tube, with both ends open.  Particles are
poured into one end and the resultant packing is allowed to settle.  In
the case where $\mu_w = 0$, no stress can be supported by the side
walls, and the granular packing is liquid-like: the particle height is
the same in both sides of the U-tube.  For $\mu_w = 0.1$, the right side
is considerably lower than the left, though some particles are higher
than the bend in the tube.  The walls support some of the stress, but
the pressure is strong enough to force some particles up the right-hand
tube.  For $\mu_w = \mu = 0.5$, there are almost no particles on the
right side of the U-tube, and all the weight on the left side is
supported by the walls of the tube.  In all these cases, the important
factor is the particle-wall friction, as the particle-particle friction
remains unchanged.

\section{\label{sec:stresses}Stresses}

Of particular interest in the construction of silos is the distribution
of stresses~\cite{Nedderman1992}.  In a liquid, hydrostatic pressure
increases with depth.  Granular materials support shear, so the side
walls of a container can support some of this shear, provided $\mu_w >
0$.  The problem of the resultant vertical stress in a silo after
filling has a long history, beginning with Janssen in 1895.  Janssen's
analysis\cite{Janssen1895,Rayleigh1906} is still in use today, although
it rests on a series of assumptions which have not been fully tested.

For a container with static wall friction
$\mu_w$ and granular pack of total height $z_0$, the Janssen analysis
predicts the vertical stress $\sigma_{zz}(z)$ at a height $z$ is
\begin{equation}
\sigma_{zz}(z) = \rho g l\left[1 -
\exp\left(-\frac{z_0 - z}{l}\right)\right]
\label{eq:janssen}
\end{equation}
where $\rho$ is the volumetric density, $l$ is the decay length, and
$z_0$ is the top of the packing.  The decay length $l$ is determined by
the geometry.  For a $2D$ container of length $L$, $l_{2D} = \frac{L}{2
\kappa \mu_w}$.  For a quasi-$2D$ container of length $L$ and width $W$,
$l_{quasi-2D} = \frac{LW}{2L + 2W}\frac{1}{\kappa\mu_w}$.  For a $3D$
cylindrical container of radius $R$, $l_{3D} =
\frac{R}{2\kappa\mu_w}$~\cite{Duran2000}.

Another two-parameter fit was proposed by Vanel and
Cl\'{e}ment~\cite{VanelMay1999} to reconcile their experimental findings
on weakly shaken packings with Janssen theory.  The fit assumes a region
of perfect hydrostaticity, followed by a region that conforms to the
Janssen theory.
\begin{eqnarray}
z_0 - z < a & : & \sigma_{zz}(z) = \rho g (z_0 - z) \nonumber\\ 
z_0 - z > a & : & \sigma_{zz}(z) = \rho g \left(a + l\left[1 - \exp\left(-\frac{z_0 - z - a}{l}\right)\right]\right)
\label{eq:twoparm}
\end{eqnarray}
The two parameters are $a$, the height of the crossover, and $l$ the
decay length, with the same values for different geometries as in the
Janssen theory.

\begin{figure}
\includegraphics[width=2.25in,angle=270,clip]{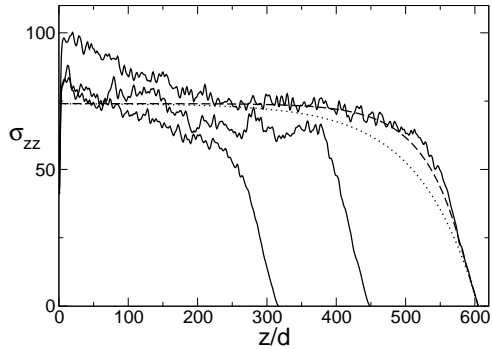}
\caption{\label{fig:stress-2d} Vertical stress $\sigma_{zz}$ for $2D$
packings.  The packings in order of decreasing depth are (a) $N \approx
20000$, length $L=30\bar{d}$, (b) $N \approx 10000$, $L=20\bar{d}$, and
(c) $N \approx 10000$, $L=30\bar{d}$.  The particles are polydisperse
with $\Delta = 0.1\bar{d}$.  The two fits to (a) are a fit to Janssen
(Eqn. 4)
(dotted line) and a fit to Vanel-Cl\'{e}ment (Eqn. 5) (dashed line).}
\end{figure}

We first present vertical stresses for the two-dimensional ($2D$) case.
These are shown in Figure~\ref{fig:stress-2d}.  Unlike previous
studies~\cite{HolstJan1999b}, these piles are extremely deep so that the
height-independent region can be easily observed.  The Janssen theory
does not describe the $2D$ stress profiles well.  There is a clear
linear hydrostatic-like region at the top of the packing that the
Janssen theory does not account for.  The Vanel-Cl\'{e}ment form is a
much closer fit to the data.  In Figure~\ref{fig:stress-2d}, the deepest
pile of length $30\bar{d}$ is fit to both forms.  The Janssen form
predicts $l/\bar{d} = 84.6$ and thus $\kappa = 0.354$, while the
Vanel-Clement form yields $a = 35$ and $l/\bar{d} = 49.6$ and thus
$\kappa = 0.605$.  As we will see below, there are a number of
differences between $2D$ and $3D$ packings.  One difference is the
height of the non-saturated region, which is on the order of $5-7 L$ in
$2D$ and $6R$ in $3D$.  There is about a factor of 2 difference between
the diameter or width of the container in $2D$ compared to $3D$.  In
addition, the saturated pressure is also much larger in $2D$ than the
$3D$ case.  Both of these effects seem to arise from the necessity of
all the force being carried by the side walls.  A similar effect is seen
at the bottom of the packing, where the slight increase in the pressure
at the bottom of the pile is over a much larger region than in the $3D$
case as shown below.  This increase occurs because geometrically the
walls can not support forces from particles below this height.

\begin{figure}
\includegraphics[width=2.25in,angle=270,clip]{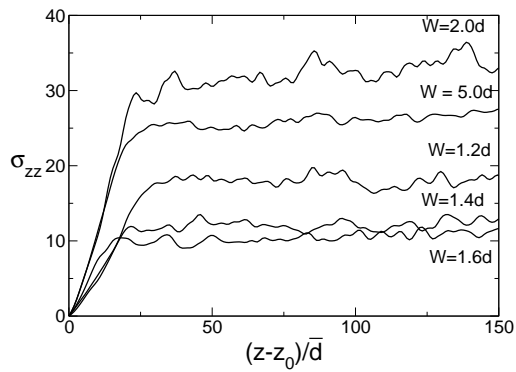}
\caption{\label{fig:quasi2d} Vertical stress for quasi-$2D$ packings
  with increasing widths, shown as $(z-z_0)/\bar{d}$, where $z_0$ is the
  top of the pile.  Six separate packings are shown, with the width $W$
  increasing from $1.2\bar{d}$ to $5.0\bar{d}$.  The
  pressure-independent height does not increase monotonically with $W$,
  and it saturates at $W=2.0\bar{d}$.  Results for Hookean springs and
  polydispersity $\Delta = 0.1\bar{d}$.}
\end{figure}

To study the crossover from $2D$ to $3D$, we studied a series of
packings in rectangular containers with increasing width, ranging from
$W = 1.2\bar{d}$ to $5\bar{d}$.  We show in Figure~\ref{fig:quasi2d} the
vertical stress profiles for these quasi-2D packings.  The stress
profiles show the same hydrostatic region followed by saturation as the
$2D$ profiles, though the hydrostatic region at the top of the pile is
much smaller in the quasi-$2D$ case.  In addition, the saturation stress
value does not increase monotonically with width.  The saturation stress
decreases as one goes from a width of $1.2\bar{d}$ to $1.4\bar{d}$, but
increases again as one reaches $2\bar{d}$.  A width of $2\bar{d}$ seems
to represent a transition to a more $3D$-like behavior, and the
saturation stress does not change appreciably for larger widths.  This
suggests that $L$ is the dominant length scale for widths greater than
$2\bar{d}$, while for smaller widths, the width $W$ is the dominant
length scale for determination of the vertical stress.

\begin{figure}
\includegraphics[width=2.25in,angle=270,clip]{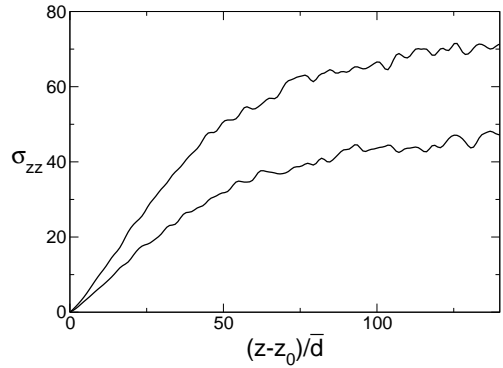}
\caption{\label{fig:quasi2d-swnf} Vertical stress for quasi-2D packings
  of length $30\bar{d}$ and width $1.6\bar{d}$ (lower curve) and width
  $2\bar{d}$ (upper curve).  The long walls of length $L=30\bar{d}$ have
  $\mu_w = 0$, while the short walls have $\mu_w = 0.5$.  The stress
  profiles are very close to those of the $2D$ simulations.}
\end{figure}

We also investigated quasi-$2D$ packings with no friction on the long
walls $\mu_w=0$, while the short side walls have $\mu_w = 0.5$.  This
system was modeled to understand how the long side walls affect the
geometry.  This system, as shown in Figure~\ref{fig:quasi2d-swnf},
produces stress profiles very different from the quasi-$2D$ case with
friction on all walls.  Two widths were used and both show much larger
final stress values and much larger hydrostatic regions.  This suggests
that both sets of walls are important for determining the stress
profiles for a wide range of widths.  The stress profiles seen in this
case resemble much more the $2D$ stress profiles presented in
Fig~\ref{fig:stress-2d}.  These results demonstrate that experiments in
thin plates meant to approximate $2D$ systems are more representative of
$3D$ systems than that of pure $2D$ systems.

\begin{figure}
\includegraphics[width=2.25in,angle=270,clip]{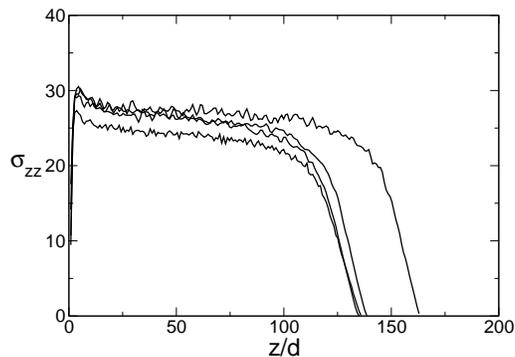}
\caption{\label{fig:stress-3d} Vertical stress $\sigma_{zz}$ for $3D$
cylindrical packings with radius $10\bar{d}$ and $N=50,000$.  The four
different packings are (a) a monodisperse ($\Delta = 0$) packing with
Hookean contacts, (b) a monodisperse packing with Hertzian contacts, (c)
a polydisperse ($\Delta/\bar{d} = 0.1$) packing with Hertzian contacts,
and (d) a polydisperse ($\Delta/\bar{d} = 0.5$) packing with Hertzian
contacts.  The packings in order of ascending height are c, a, b, d.
Results for case a are averaged over 4 runs while the other three cases
are for one run each.  Results for $\mu=\mu_w=0.5$.}
\end{figure}

We show in Figure~\ref{fig:stress-3d} vertical stress profiles for $3D$
systems in cylindrical packings.  Changing the force model from a
Hookean to a Hertzian interaction has a very small effect on the final
packing.  Changing from a monodisperse to a very highly polydisperse
system ($\Delta/\bar{d} = 0.5$) also does not change the overall stress
profile, although it does increase the height of the packing.  This
suggests that although polydispersity is required in $2D$ systems to
avoid crystallization, it has little effect on the stress profile.  It
is thus reasonable to compare monodisperse $3D$ packings to polydisperse
$2D$ systems.

In this case we see that the important transition is from the $2D$
system to the quasi-$2D$ system.  In most respects the quasi-$2D$ system
is much closer to the $3D$ system than the pure $2D$ system.  The size
of the hydrostatic region changes most markedly from the pure $2D$
system to the quasi-$2D$ system.  The stress values in the depth of the
pile are much larger for the $2D$ case than the quasi-$2D$ case and are
comparable to the $3D$ case.

\section{\label{sec:Forces}Distribution of Forces}

\begin{figure}
\includegraphics[width=2.25in,angle=270,clip]{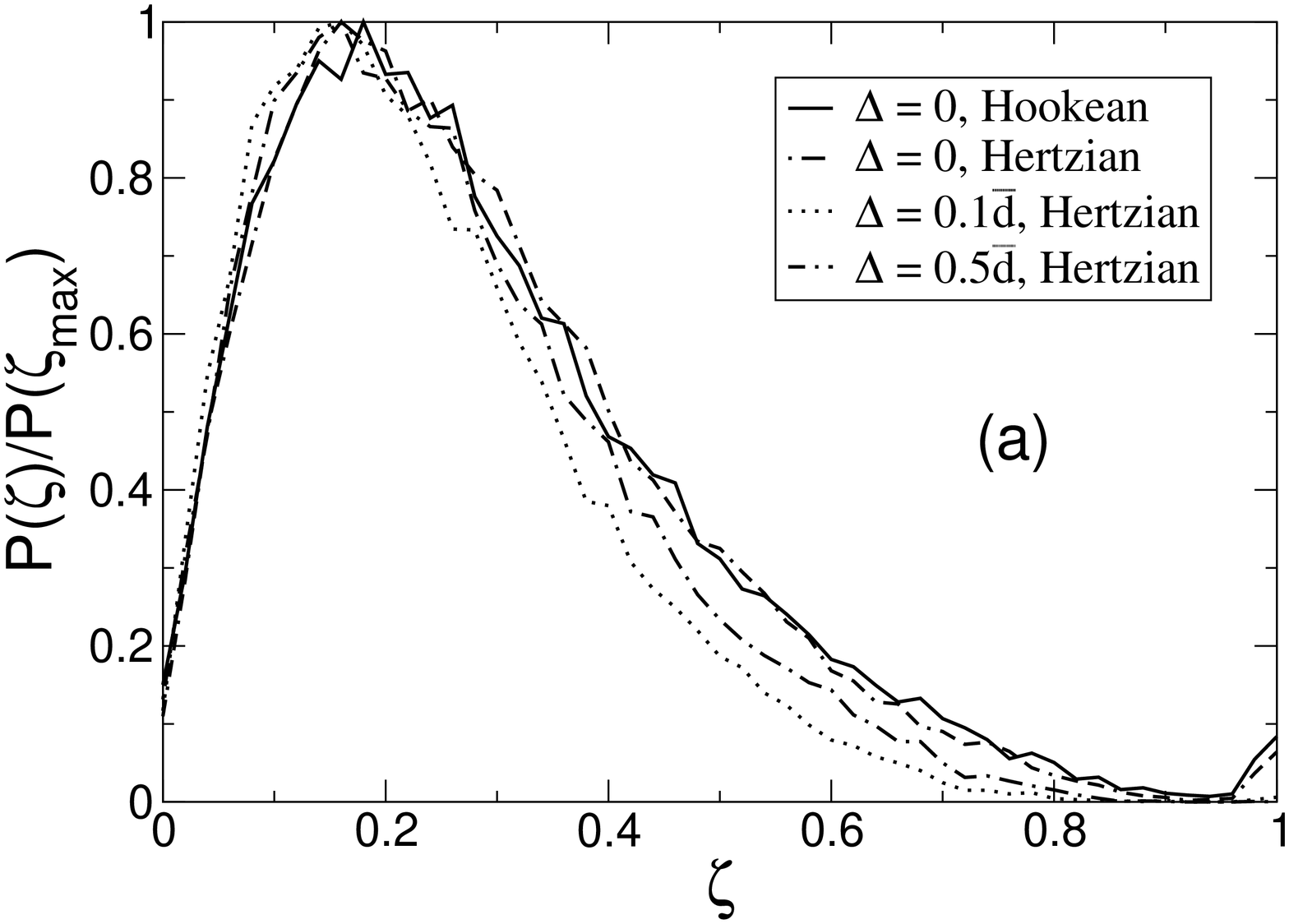}
\includegraphics[width=2.25in,angle=270,clip]{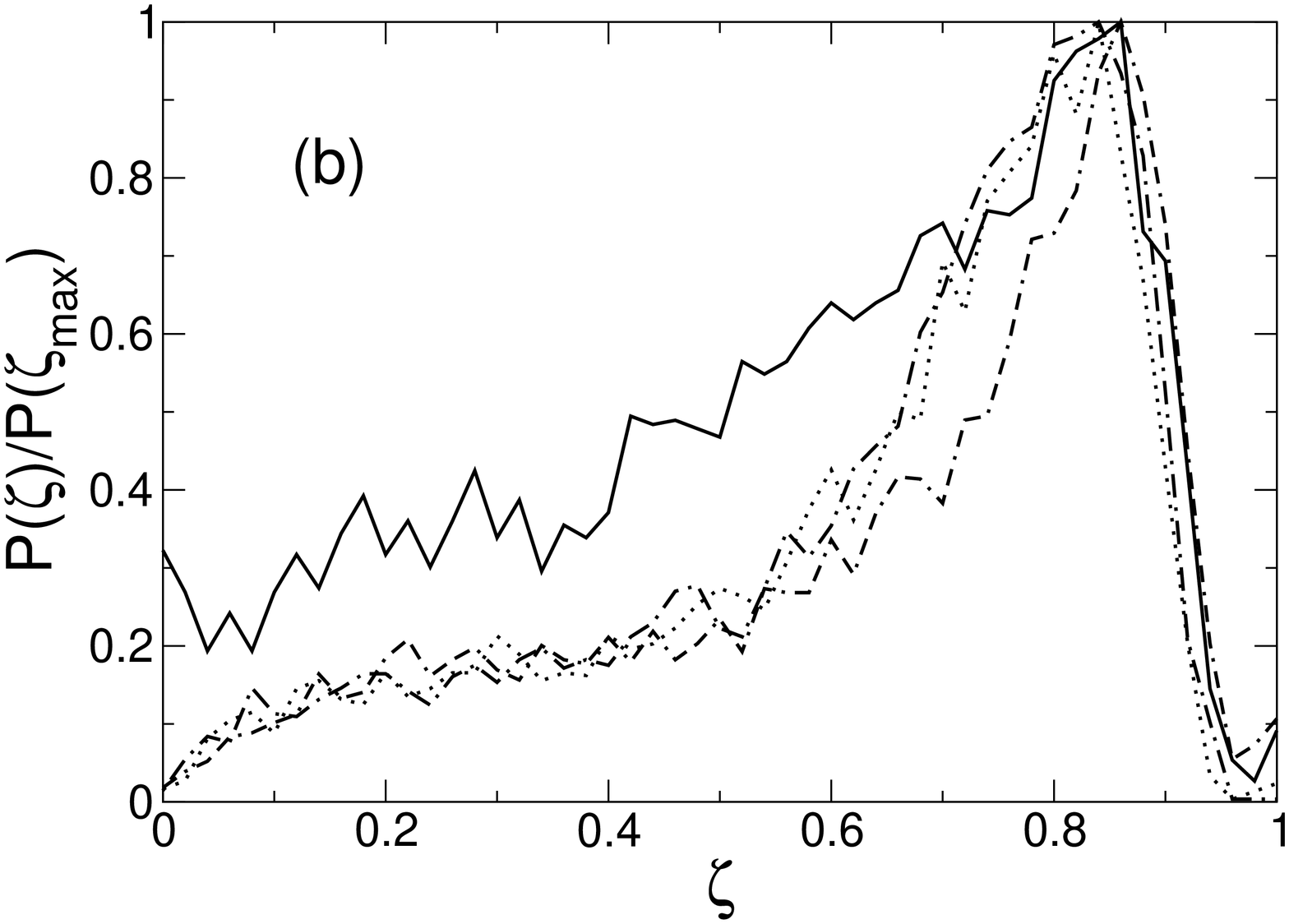}
\caption{\label{fig:coulomb}Probability distributions $P(\zeta)$ in the
height-independent pressure region in the bulk of the packing (a) and at
the side walls (b), each normalized by its maximum value
$P(\zeta_{\mathrm{max}})$.  $\zeta = F_t/\mu F_n$ in (a) and $\zeta =
F_t/\mu_w F_n$ in (b).  The legend applies to both figures.  Forces in
the bulk are far from the Coulomb failure criterion, while those at the
walls are very close to it.}
\end{figure}

The previous section has demonstrated that the stress profiles of our
granular packings do not conform to the Janssen theory.  The historical
assumption has been that the assumption of incipient failure in the
Janssen analysis is responsible for the failure of the theory.  We thus
tested directly the Janssen assumptions of incipient failure in our
simulations, i.e. whether the tangential forces at the wall are actually
at the Coulomb yield criterion $F_t = \mu_w F_n$.  We define $\zeta =
F_t/\mu F_n$ in the bulk of the packing and $\zeta = F_t/\mu_w F_n$ for
forces at the wall.  If a specific force is at the Coulomb failure
criterion, $\zeta = 1$.  We analyzed the $3D$ systems shown in
Figure~\ref{fig:stress-3d}.  By examining the distribution of forces in
the interior of our packings, we find that no particle-particle contacts
are at the Coulomb criterion irrespective of method or level of
polydispersity, as shown in Figure~\ref{fig:coulomb}a.  We find similar
results in the interior of $2D$ packings.  However the particle-wall
forces in the height-independent stress region are much closer to the
Coulomb criterion.  In the four systems examined, the bulk of the
particle-wall tangent forces are close to incipient failure, in contrast
to the particle-particle tangent forces in the bulk, as seen in
Figure~\ref{fig:coulomb}b.  The amount of polydispersity and the contact
force model have essentially no effect on how close the system is to
incipient failure.  We have also investigated the role of $\mu_w > \mu$
in Ref.~\cite{LandryApr2003}.  In this case, the particle-wall forces
are always less than the Coulomb criterion, as the walls cannot support
larger tangential forces than those supported in the bulk.

\begin{figure}
\includegraphics[width=2.25in,angle=270,clip]{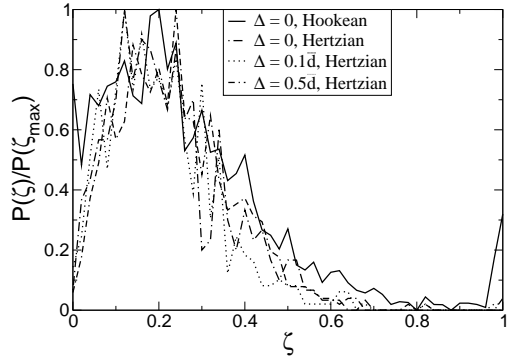}
\caption{\label{fig:coulomb-top}Probability distributions $P(\zeta)$ at
the side wall in the linear hydrostatic region at the top of the packing
with $\mu = \mu_w = 0.5$, each normalized by its maximum value
$P(\zeta_{\mathrm{max}})$.  $\zeta = F_t/\mu_w F_n$.  In contrast to the
behavior in the height-independent pressure region, the forces at the
walls are far from the Coulomb failure criterion in all cases.}
\end{figure}

As discussed above, all of our packings exhibit a linear region in the
stress profile at the top of the packing.  We can also examine the
Janssen assumption in this linear region.  In contrast to the situation
in the height-independent stress region of the packing, near the top of
the pile the forces at the wall are far from the Coulomb criterion, as
shown in Figure~\ref{fig:coulomb-top}.  This explains why the Janssen
form is not obeyed in this region: the walls in this region do not
support stress and thus the stress profile is hydrostatic.

\section{\label{sec:conclusion}Conclusions}

We examined the differences between vertical stress in granular packings
in $2D$, quasi-$2D$, and $3D$.  We focused on the vertical stress of these
packings and observed that $2D$ packings support much less vertical
stress than $3D$ packings.  In contrast, quasi-$2D$ packings (those
packings between thin walls) rapidly converge to the behavior of $3D$
packings as the wall thickness is increased, though their progression is
not monotonic.  We also show that polydispersity and different force
models do not have a large effect on the vertical stress in $3D$
packings.  In all cases, we show that the Janssen assumption of
incipient failure is not justified everywhere in the packing.

We acknowledge discussions with L. Silbert and Jin Ooi, and thank
L. Silbert for his critical reading of the manuscript.  This work was
supported by the Division of Materials Science and Engineering, Basic
Energy Sciences, Office of Science, U.S. Department of Energy.  This
collaboration was performed under the auspices of the DOE Center of
Excellence for the Synthesis and Processing of Advanced Materials.
Sandia is a multiprogram laboratory operated by Sandia Corporation, a
Lockheed Martin Company, for the United States Department of Energy's
National Nuclear Security Administration under contract
DE-AC04-94AL85000.

\bibliography{grain}

\end{document}